\documentclass[preprint, showpacs,preprintnumbers,amsmath,amssymb,nofootinbib]{revtex4}
\usepackage{mathrsfs}
\usepackage{amsmath}
\usepackage{amsfonts}
\usepackage{latexsym}
\usepackage{amsfonts}
\usepackage{graphicx}
\usepackage{epsf}
\usepackage{dcolumn}
\usepackage{bm}

\newcommand{\bea}[1]{\begin{eqnarray}\label{#1}}
\newcommand{\eea}{\end{eqnarray}}

\def\gsim{ \lower .75ex \hbox{$\sim$} \llap{\raise .27ex \hbox{$>$}} }
\def\lsim{ \lower .75ex \hbox{$\sim$} \llap{\raise .27ex \hbox{$<$}} }

\begin{document}
 \title{The stability of the Einstein static state in $f(T)$ gravity}

\author{Puxun Wu and  Hongwei Yu }
\address
{Center of Nonlinear Science and Department of Physics, Ningbo
University,  Ningbo, Zhejiang, 315211 China }

\begin{abstract}
The stability of the Einstein static universe against the
homogeneous scalar perturbations in $f(T)$ gravity is analyzed. Both
the spatial closed  and  open universes  are considered. We find
that  the stable Einstein static solutions exist in both cases.
Considering a concrete $f(T)$ model and assuming that the cosmic
energy  has a constant equation of state $w$, we obtain that, in the
closed case, $w<1/3$ is required. Thus, $f(T)$ theory gives a larger
region of $w$ than that in general relativity ($-1<w<-1/3$) to have
the stable Einstein static solution. For the open universe, $f(T)$
theory allows the stable Einstein static solution,  although this kind of solution is
forbidden in general relativity. Thus, a modification of gravity can
play a crucial role in stabilizing the Einstein static solution.

\end{abstract}
\maketitle

\section{Introduction}\label{sec1}
One of the most important and challenging problems in modern
cosmology is how to explain the current cosmic acceleration,
discovered firstly from the type Ia supernova
observations~\cite{Riess1998} and then further confirmed by the
cosmic microwave background radiation~\cite{Spergel2003} and the
large scale structures~\cite{Tegmark2004}. One  popular way to
explain this observed phenomenon is to postulate, within the context
of general relativity, the existence of an exotic energy component,
called dark energy (see \cite{DE} for recent review), in our
universe. Another is to modify the general relativity.  $f(R)$
theory~(see \cite{Felice2010} for recent review), obtained by
generalizing the Ricci scalar  $R$ in the Einstein-Hilbert action to
an arbitrary function $f$ of $R$, is one of such modified gravity
theories

In 1928, Einstein~\cite{Einstein1928} first introduced the
teleparallel gravity (TG) in his endeavor to unify gravity and
electromagnetism with the introduction of a tetrad field. Although
not succeeding, as is well known,  TG can, however, show up as a
theory completely equivalent to general
relativity~\cite{Einstein1930, Moller1961}. Since TG is built on the
teleparallel geometry, and  the Weitzenb\"{o}ck connection rather
than the Levi-Civita connection is used in this geometry, the
Riemann curvature vanishes automatically and the spacetime has only
torsion. The torsion scalar $T$ is the Lagrangian density of TG.

Recently, in analogy to  $f(R)$ theory, a new modified gravity to
account for the accelerating  cosmic expansion, named $f(T)$ theory,
is proposed by extending the TG action $T$ to an arbitrary function
$f$ of $T$.  $f(T)$ theory can not only  explain the present cosmic
acceleration with no need of dark energy~\cite{Bengochea2009}, but
also provide an alternative to inflation without an
inflaton~\cite{Ferraro2007, Ferraro2008}. Moreover, further studies
have shown that  $f(T)$ may avoid the big bang singularity problem
in the standard cosmology~\cite{Ferraro2011}, realize the crossing
of phantom divide line for the effective equation of
state~\cite{Wu2011, Bamba2011},  fit the current type Ia supernova
observation very well~\cite{Wu2010a} and yield an usual early cosmic
evolution~\cite{Wu2010b}. It, thus, has recently spurred an
increasing deal of interest~\cite{FT, Ferraro2011a, Li2011,
Miao2011}. It is worth pointing out here that $f(T)$ gravity also
suffers from some problems, such as the violation of local Lorentz
invariance~\cite{Li2011} and the violation of the first law of black
hole thermodynamics~\cite{Miao2011}.

In this paper, we plan to analyze the stability of the Einstein
static state in $f(T)$ theory.  Both spatially closed and open
universes are considered. Our interest in this issue lies in  that
our universe might have originated from the Einstein static state
and then evolved to the inflation, so as to provide a possible way
to resolve the big bang singularity problem~\cite{Ellis2004a,
Ellis2004b}. The Einstein static universe has attracted  a great
deal of attention~\cite{Carneiro2009, Mulryne2005, Parisi2007,
Wu2009, Lidsey2006, Bohmer2007, Seahra2009, Bohmer2009, Barrow2003,
Barrow2009, Goswami2008, Clifton2005,Boehmer2010, Boehmer20093,
Wu20092, Odrzywolek2009, Goheer2009, Canonico2010, Zhang2010}. For
instance, the Einstein static solutions have been analyzed  in
braneworld theory, Hoava-Lifshitz gravity and loop quantum
cosmology~\cite{Mulryne2005, Parisi2007, Wu2009, Boehmer20093,
Wu20092, Lidsey2006}. The stability of the Einstein static state has
been studied in $f(R)$ gravity~\cite{Bohmer2007,Goswami2008,
Seahra2009,Goheer2009} and it was found that, in several concrete
$f(R)$ models, the stable solutions do exist under the homogeneous
perturbations~\cite{Bohmer2007}. However, Goswami et al.
\cite{Goswami2008} argued that only one functional form of $f(R)$
admits an Einstein static solution, which seems to be inconsistent
with what was obtained in \cite{Bohmer2007}. This contradiction was
reconciled in~\cite{Seahra2009} by considering the stability of the
Einstein static universe under the homogeneous and inhomogeneous
scalar perturbations in a general $f(R)$ theory. It is worth noting
that all above studies are done in the case of a spatially closed
universe. More recently, it was found that in the frameworks of
Loop Quantum Cosmology and Horava-Lifshitz gravity,  the Einstein
static solution may also exist in an open
universe~\cite{Canonico2010}.

We examine, in the present paper, the stability of the Einstein
static solution against homogeneous perturbations in $f(T)$ gravity.
In the following section, we give a brief review on $f(T)$ gravity.
In section III, we give the Einstein static solution and then
discuss its stability by analyzing the homogeneous scalar
perturbations, and we conclude in Section IV.

\section{ the $f(T)$ theory}
In this section, we  give a brief review on $f(T)$ gravity. In this
theory, the dynamical object is the vierbein $e^\mu_i$ rather than
the metric. If $e_\mu^i$  is the inverse matrix of vierbein
$e^\mu_i$, the relation between them is
\begin{eqnarray}\label{Eq1}
e^\mu_i e_\mu^j=\delta^j_i, \quad e^\mu_i e_\nu^i=\delta^\mu_\nu,
\end{eqnarray}
where  $i$ is an index running over $0,1,2,3$ for the tangent space
of the manifold, and $\mu$, also running over   $0,1,2,3$, is the
coordinate index on the manifold.  This vierbein relates with the
metric through
\begin{eqnarray}\label{gu1} g_{\mu\nu}=\eta_{ij}e^i_\mu e^j_\nu\;,
\end{eqnarray}
where $\eta_{ij}=diag(1,-1,-1,-1)$. Using Eq.~(\ref{Eq1}), the above
expression can be inverted to obtain
\begin{eqnarray}
\eta_{ij}=g_{\mu\nu}e_i^\mu e_j^\nu\;,\end{eqnarray} which means
that the vierbein is orthonormal.

As already mentioned in the previous section,  TG uses the
curvatureless Weitzenb\"{o}ck connection, which is defined as
\begin{eqnarray}  {{\Gamma}}^\lambda_{\mu\nu}=e^\lambda_i\partial_\nu e^i_\mu=- e^i_\mu \partial_\nu e^\lambda_i\;.
\end{eqnarray}
From this connection, one can introduce a non-null torsion tensor
$T^\sigma_{\;\;\mu\nu}$,
\begin{eqnarray}
T^\sigma_{\;\;\mu\nu}={{\Gamma}}^\sigma_{\nu\mu}-{{\Gamma}}^\sigma_{\mu\nu}
\;.
\end{eqnarray}
Defining other two tensors:
\begin{eqnarray}S^{\;\;\mu\nu}_\sigma\equiv\frac{1}{2}(K^{\mu\nu}_{\;\;\;\;\sigma}+\delta^\mu_\sigma
T^{\alpha \nu}_{\;\;\;\;\alpha}-\delta^\nu_\sigma T^{\alpha
\mu}_{\;\;\;\;\alpha})\;,
\end{eqnarray}
 and
\begin{eqnarray}
 K^{\mu\nu}_{\;\;\;\;\sigma}=-\frac{1}{2}(T^{\mu\nu}_{\;\;\;\;\sigma}-T^{\nu\mu}_{\;\;\;\;\sigma}-T_{\sigma}^{\;\;\mu\nu}),
 \end{eqnarray} one can obtain the torsion scalar $T$
\begin{eqnarray}\label{ST}
T\equiv S^{\;\;\mu\nu}_\sigma T^\sigma_{\;\;\mu\nu}\;,
 \end{eqnarray}
which is the  teleparallel Lagrangian. Thus the TG action can be
expressed as \begin{eqnarray} I=\frac{1}{16\pi G}\int d^4 x\; e\;
T\;,\end{eqnarray} where $e=\det(e_\mu^i)=\sqrt{-g}$. Since  $eT$
just differs from the Einstein-Hilbert Lagrangian $eR$ by a total
derivative term \begin{eqnarray}\label{BT1}
-e\;R=e\;T-2\partial_\nu(e\;T_\mu^{\;\;\mu\nu})\;\end{eqnarray}
where $R$ is the scalar curvature for the Levi-Civita connection, TG
is completely equivalent to general relativity.

As $f(R)$ gravity, the action of $f (T)$ theory is obtained by
replacing $T$ in the TG action by a general function of  $T$
 \begin{eqnarray} I=\frac{1}{16\pi G}\int
d^4 x\; e\; f(T)\;.\end{eqnarray} Apparently the last term in
Eq.~(\ref{BT1}) can be discarded by converting  it to a boundary
term in TG.  It,  however, remains in $f(T)$ theory, which leads to
the violation of local Lorentz invariance. Adding a matter term in
the above equation and doing a derivative with respect to vierbein,
one can obtain the field equation of  $f(T)$ gravity.

\section{the Einstein static universe in  $f(T)$ theory}

To analyze the Einstein static universe, we consider the FRW
universe with non flat spatial sections. Due to the lack of local
Lorentz invariance,  pairs of vierbein fields connected by local
Lorentz transformations are inequivalent. Thus, one should be
careful in obtaining the parallelized frames.
Here, we follow the procedure given in
Ref.~\cite{Ferraro2011a} to get the vierbein.  For a closed universe,  the vierbein is
\begin{eqnarray}\label{v1}
e^0=dt\;;\quad e^1=a(t)E^1\;;\quad e^2=a(t)E^2\;;\quad e^3=a(t)E^3\;,
\end{eqnarray}
with
\begin{eqnarray}
&&E^1= -\cos \theta d \psi + \sin \psi \sin \theta (\cos \psi d\theta-\sin \psi \sin \theta d\phi)\\ \nonumber
&&E^2 = \sin \theta \cos \phi d\psi-\sin \psi[(\sin \psi \sin \phi- \cos \psi \cos \theta \cos \phi) d\theta\\ \nonumber &&\qquad\qquad+ (\cos\psi\sin \phi + \sin\psi\cos\theta\cos\phi) \sin\theta d\phi]\\ \nonumber
&&E^3 = -\sin \theta \sin\phi d\psi -\sin\psi[(\sin\psi \cos\phi + \cos\psi\cos\theta\sin\phi)d\theta\\ \nonumber &&\qquad\qquad +(\cos\psi\cos\phi- \sin\psi \cos\theta \sin\phi) \sin\theta d\phi],
\end{eqnarray}
and, for an open one, it is
\begin{eqnarray}\label{v2}
e^0=dt\;;\quad e^1=a(t)\bar{E}^1\;;\quad e^2=a(t)\bar{E}^2\;;\quad e^3=a(t)\bar{E}^3\;,
\end{eqnarray}
with
\begin{eqnarray}
&&\bar{E}^1 = \cos\theta d\psi +\sinh\psi\sin\theta(-\cosh\psi d\theta + i \sinh\psi \sin\theta d\phi)\\ \nonumber
&&\bar{E}^2 = -\sin\theta \cos\phi d\psi  + \sinh\psi [(i \sinh\psi \sin\phi - \cosh\psi \cos\theta \cos\phi) d\theta \\ \nonumber && \qquad\qquad+ (\cosh\psi\sin\phi + i \sinh\psi\cos\theta \cos\phi) \sin\theta d\phi]\\ \nonumber
&&\bar{E}^3 = \sin\theta\sin\phi d\psi + \sinh\psi [(i \sinh\psi\cos\phi + \cosh\psi \cos\theta\sin\phi) d\theta \\ \nonumber && \qquad\qquad + (\cosh\psi \cos\phi - i \sinh\psi\cos\theta\sin\phi) \sin\theta d\phi].
\end{eqnarray}
Here, the angular coordinates range in the intervals $0 \leq \phi\leq 2\pi$, $0 \leq \theta \leq \pi$ and $0 \leq \psi  \leq \pi$. Thus, using Eqs.~(\ref{gu1}, \ref{v1}, \ref{v2}), one can obtain the induced metric
\begin{eqnarray}
ds^2=dt^2-k^2a^2(t)[d(k\psi)^2+\sin^2(k\psi)(d\theta^2+\sin^2\theta
d\phi^2)]\;,
\end{eqnarray}
where $k=1$ for the closed universe and $k=i$ for the open universe. The torsion scalar can be expressed as
\begin{eqnarray} T=6(\pm a^{-2}-H^2)\;,\end{eqnarray} where $+$ and
$-$ correspond to the closed and open cases respectively, and
$H=\frac{\dot{a}}{a}$ is the Hubble parameter. In a non flat
universe, the modified Friedmann equation can be expressed
as~\cite{Ferraro2011}
\begin{eqnarray}\label{F1}
12H^2f'(T)+f(T)=16\pi G \rho\equiv \kappa \rho\;,
\end{eqnarray}
\begin{eqnarray}\label{F2}
4(\pm
a^{-2}+\dot{H})(12H^2f''(T)+f'(T))-f(T)-4f'(T)(2\dot{H}+3H^2)=\kappa
p\;,\end{eqnarray} where $f'(T)=df/dT$, $f''(T)=d^2f/dT^2$ and we
assume that the matter-content of the universe is a perfect fluid with
$\rho$ and $p$ being the unperturbed energy density and pressure
respectively, which satisfy the conservation equation
\begin{eqnarray}
\dot{\rho}+3H(\rho+p)=\dot{\rho}+3H(1+w)\rho=0\;.\end{eqnarray} Here
we let $w=\frac{p}{\rho}$ be a constant.

For the  Einstein static universe, we have \begin{eqnarray}
a=a_0=const,\quad \dot{a}=H=0\;,\end{eqnarray}
$T_0=T(a_0)=\frac{6}{a^2}$ for the closed universe, and
$T_0=-\frac{6}{a^2}$ for the open one. Thus, using Eqs.~(\ref{F1},
\ref{F2}) we obtain the conditions for the existence of an Einstein
static universe
\begin{eqnarray}\label{ES1}f_0=f(T_0)=\kappa \rho_0\;, \quad
\pm\frac{4f'_0}{a_0^2}-f_0=\kappa p_0\;,\end{eqnarray} with
$f'_0\equiv \frac{df}{dT}|_{T=T_0}$, $\rho_0=\rho(a_0)$ and
$p_0=p(a_0)$. Combining the above two expressions and using $T_0$,
one finds
\begin{eqnarray}\label{C1}\bigg(\frac{Tf'}{f}\bigg)_{T=T_0}=\frac{3}{2}(1+w)\;,\end{eqnarray}
which gives a constraint on $f(T)$ to obtain the Einstein static
solution.

Now we consider the stability of the Einstein static solution
against the linear homogeneous scalar perturbations. Thus, the
perturbations in the cosmic scale factor and in the energy density
 depend only on time and can be expressed as
\begin{eqnarray}
a(t)=a_0(1+\delta a(t))\;, \quad  \rho(t)=\rho_0(1+\delta
\rho(t))\;.
\end{eqnarray}
Substituting the above equation into Eqs.~(\ref{F1}, \ref{F2}) and
linearizing the results, we have \begin{eqnarray}f'_0\delta
T=\kappa\rho_0\delta\rho\;,\end{eqnarray}
\begin{eqnarray}4(\mp 2a_0^{-2}\delta a + \delta\ddot{ a})f'_0\pm 4a_0^{-2}f''_0\delta T-f'_0\delta T-8 f'_0 \delta\ddot{ a}=\kappa\delta p\;, \end{eqnarray}
where $\delta f=f'_0\delta T$ and $\delta f'=f''_0\delta T$.  Using
$\delta T=-2 T_0 \delta a$, $\delta p= w \rho_0\delta\rho$ and
$a_0^{-2}=\pm \frac{f_0}{4f'_0}(1+w)$, one can obtain
\begin{eqnarray}
\delta \ddot{a}=\frac{\kappa\rho_0}{4{f'_0}^2}(1+w)\bigg( (1+3w)
f'_0-3 (1+w)\frac{f_0f''_0}{f'_0}\bigg)\delta a\;,\end{eqnarray}
which admits a solution
\begin{eqnarray} \delta a(t)=C_1 e^{\omega
t}+C_2 e^{-\omega t}\;,
\end{eqnarray}
where $C_1$ and $C_2$ are integration constants, and $\omega$ is
given by
\begin{eqnarray}\label{O1}
\omega^2=\frac{\kappa\rho_0}{4{f'_0}^2}(1+w)\bigg((1+3w) f'_0-3
(1+w) \frac{f_0f''_0}{f'_0}\bigg)\end{eqnarray} Obviously, if
$\omega^2<0$ we have oscillating perturbation modes, which
correspond to  existence of the stable Einstein static universe.

In what follows, we divide our discussion into the closed and open universes respectively,
and consider a concrete power law $f(T)$ model in order to give the detail
conditions for  the stable Einstein static solution.
\subsection{The closed universe}
In this case, $T_0=6a_0^{-2}$ for the Einstein static solution. We
consider a model
\begin{eqnarray}\label{M1}f(T)=T+\alpha\frac{a_0^4}{6}T^2-\Lambda\;,\end{eqnarray}
where $\alpha$ is a constant and $\Lambda$ is the cosmological
constant. Substituting the above expression into the Einstein static
solution given in Eq.~(\ref{ES1}), one can obtain
\begin{eqnarray}\label{L1}\Lambda=\frac{1}{2} (-12 \alpha + \kappa\rho_0(1 + 3
w))\;,\quad \frac{1}{a_0^2}=\frac{1}{2}(-8 \alpha + \kappa
\rho_0(1+w)).\end{eqnarray} Using Eq.~(\ref{L1}), it is easy to
prove that the model (Eq.~(\ref{M1})) satisfies the constraints
given in Eq.~(\ref{C1}). A positive $a_0^2$ gives a constraint on
$\alpha$:
\begin{eqnarray}\alpha < \frac{\kappa
\rho_0}{8}(1+w).\end{eqnarray} Substituting Eq.~(\ref{M1}) into
Eq.~(\ref{O1}), we have
\begin{eqnarray}\label{O2}\omega^2 &=& \frac{\kappa\rho_0}{4 {f'_0}^2}
\frac{ (1 + w)[-16 \alpha+ \kappa \rho_0 (1+w)(1 + 3 w)]}{-8 \alpha
+ \kappa \rho_0 (1 + w)}\nonumber \\
&=&\frac{\kappa\rho_0 a_0^2}{8 {f'_0}^2} (1 + w)[-16 \alpha+ \kappa
\rho_0 (1+w)(1 + 3 w)]\;.\end{eqnarray}
 As expected,  the limit
$\alpha\rightarrow 0$ ($f\rightarrow T-\Lambda$) yields
\begin{eqnarray}
\omega^2=\frac{\kappa\rho_0}{4}(1+w)(1+3w)\;, \quad
\Lambda=\frac{\kappa\rho_0}{2} (1 + 3 w)\;,\quad \frac{1}{a_0^2}=\frac{\kappa
\rho_0}{2}(1+w)\;,\end{eqnarray} which is
just the general relativistic result. Apparently, within the general
relativity context  the solution is stable in the region
\begin{eqnarray}
-1<w<-1/3 \;,
\end{eqnarray} which violates the strong energy condition and leads
to a negative cosmological constant.

From Eq.~(\ref{O2}), we find
two stable regions in $f(T)$ gravity:
\begin{eqnarray}w<-1\;, \quad \alpha < \frac{\kappa
\rho_0}{8}(1+w)\;, \end{eqnarray} and \begin{eqnarray}
 -1<w<1/3\;, \quad \frac{\kappa \rho_0}{16}(1+w)
(1+3w)<\alpha < \frac{\kappa \rho_0}{8}(1+w)\;.
\end{eqnarray}
 Apparently, $f(T)$ gravity enlarges the allowed region
of $w$ for obtaining the stable Einstein static solution, and the
strong energy condition is not necessarily  violated. In addition,
different from the general relativity case where the cosmological
constat is negative at the  static point,  $\Lambda$ can be either
positive or negative in $f(T)$ theory as shown in Tab.~(\ref{Tab1}).

\begin{table}[!h]
\tabcolsep 0pt \caption{The property of $\Lambda$ for the stable
Einstein static solution in a closed universe.} \vspace*{-12pt}
\begin{center}\label{Tab1}
\def\temptablewidth{0.8\textwidth}
{\rule{\temptablewidth}{1pt}}
\begin{tabular*}{\temptablewidth}{@{\extracolsep{\fill}}ccc}
&$\alpha$ &$\Lambda$ \\   \hline $w<-1$  & $\frac{\kappa
\rho_0}{12}(1+3w)<\alpha<\frac{\kappa
\rho_0}{8}(1+w)$     & $\Lambda>0$  \\
      & $\alpha<\frac{\kappa \rho_0}{12}(1+3w)$     &
$\Lambda<0$ \\
        $-1<w<-1/3$        &  $\frac{\kappa \rho_0}{16}(1+w)(1+3w)<\alpha<\frac{\kappa
\rho_0}{8}(1+w)$     & $\Lambda>0$  \\
    $-1/3<w<1/3$  & $\frac{\kappa \rho_0}{16}(1+w)(1+3w)<\alpha <\frac{\kappa
\rho_0}{12}(1+3w)$     & $\Lambda<0$  \\
   & $\frac{\kappa \rho_0}{12}(1+3w)<\alpha <\frac{\kappa
\rho_0}{8}(1+w)$     & $\Lambda>0$  \\
       \end{tabular*}
       {\rule{\temptablewidth}{1pt}}
       \end{center}
       \end{table}

\subsection{The open universe}
This case corresponds to $T_0=-6a_0^{-2}$ in the Einstein static
universe. We consider the same model as  that in the closed case.
 Substituting Eq.~(\ref{M1}) into the Einstein static
solution given in Eq.~(\ref{ES1}), we have
\begin{eqnarray}\Lambda=\frac{1}{2} (-12 \alpha + \kappa\rho_0(1 + 3
w))\;,\quad \frac{1}{a_0^2}=\frac{1}{2}(8 \alpha - \kappa
\rho_0(1+w)).\end{eqnarray} A positive $a_0^2$ leads to $\alpha$:
\begin{eqnarray}\alpha > \frac{\kappa
\rho_0}{8}(1+w).\end{eqnarray} Doing the same calculation as in the
closed case, we obtain
\begin{eqnarray}\label{O3}\omega^2&=&\frac{\kappa\rho_0}{4 {f'_0}^2}\frac{(1 +
w)[-16\alpha+\kappa\rho_0(1+w)(1+3w)}{-8\alpha+\kappa \rho_0 (1 +
w)}\nonumber \\ &=&\frac{\kappa\rho_0 a_0^2}{8
{f'_0}^2}(1+w)[16\alpha-\kappa\rho_0(1 + w)(1+3w)]\;.\end{eqnarray}

When $\alpha=0$, we  get
\begin{eqnarray} \omega^2=\frac{\kappa\rho_0}{4}(1+w)(1+3w)\;, \quad
\Lambda=\frac{\kappa\rho_0}{2} (1 + 3 w)\;,\quad
\frac{1}{a_0^2}=-\frac{\kappa \rho_0}{2}(1+w)\;.\end{eqnarray} The
stable solution requires $-1<w<-1/3$, but it leads to $a_0^2<0$.
Thus, in the spatially open case, the general relativity does not
allow a stable Einstein static solution.

In $f(T)$ theory, from Eq.~(\ref{O3}), we find that the conditions
for stability are
\begin{eqnarray} w<-1\;, \quad \alpha > \frac{\kappa
\rho_0}{16}(1+w)(1+3w)\;, \end{eqnarray} and
\begin{eqnarray} w>1/3\;, \quad \frac{\kappa
\rho_0}{8}(1+w) <\alpha < \frac{\kappa \rho_0}{16}(1+w)(1+3w) \;,
\end{eqnarray} which means that a phantom or a stiff matter is required for
obtaining the stable Einstein static universe. In Tab.~(\ref{Tab2}),
we show the properties of $\Lambda$.

\begin{table}[!h]
\tabcolsep 0pt \caption{The property of $\Lambda$ for the stable
Einstein static solution in an open universe.} \vspace*{-12pt}
\begin{center}\label{Tab2}
\def\temptablewidth{0.8\textwidth}
{\rule{\temptablewidth}{1pt}}
\begin{tabular*}{\temptablewidth}{@{\extracolsep{\fill}}ccc}
&$\alpha$ &$\Lambda$ \\   \hline $w<-1$  & $\alpha>\frac{\kappa
\rho_0}{16}(1+w)(1+3w)$     &
$\Lambda<0$ \\
        $w>1/3$        &  $\frac{\kappa \rho_0}{12}(1+3w)<\alpha<\frac{\kappa
\rho_0}{16}(1+w)(1+3w)$     & $\Lambda<0$  \\
       & $\frac{\kappa \rho_0}{8}(1+w)<\alpha <\frac{\kappa
\rho_0}{12}(1+3w)$     & $\Lambda>0$  \\
       \end{tabular*}
       {\rule{\temptablewidth}{1pt}}
       \end{center}
       \end{table}


\section{Conclusion}
The  Einstein static universe has been proposed as the asymptotic
origin of our universe for avoiding the big bang singularity
problem. Thus, in order to assure that the universe can stay at this
static state past-eternally, a stable Einstein static universe is
essential. In this paper, we have  analyzed the stability of the
Einstein static universe against the homogeneous scalar
perturbations in $f(T)$ gravity, which is a new modified gravity
used to account for the present accelerated cosmic expansion and
explain the comic inflation. Different from usual discussions
considering  only the spatially closed universe, both the closed and
open cases are studied in the present paper. In particular, we
assume that the matter-content of the universe is a perfect fluid and it has a
constant equation of state $w$. By considering a concrete pow law
$f(T)$ model: $f(T)=T+\alpha\frac{a_0^4}{6}T^2-\Lambda$, we find
that, in the closed case, $w<1/3$ is required to obtain a stable
Einstein static solution, which means that the strong energy
condition violated in general  relativity can be satisfied.  Our
results show a larger allowed region of $w$ in $f(T)$ gravity than
that  in general relativity where the  condition for stability is
$-1<w<-1/3$. As shown in Tab.~(\ref{Tab1}), $f(T)$ theory allows
both negative and positive $\Lambda$ in contrast to  general
relativity in which  a negative cosmological constant is needed.

For the open universe, we find that there is no  stable Einstein
static solution in general relativity, but $f(T)$ theory allows it.
The stable Einstein static universe requires that the perfect fluid
is a phantom ($w<-1$) or a stiff matter ($w>1/3$). Therefore, we can
conclude that the modification of gravity can play a crucial role in
stabilizing the Einstein static solution.

We only consider the power law $f(T)$ model in the present paper.
However, one can show that there also exists the Einstein static
solution in the exponential model, but the stability conditions are
much more complicated than that in the pow law case. So, we do not
give the details here. Finally, we must point out that here only the
homogeneous perturbations are analyzed. It is of interest to extend
our results to the case of the inhomogeneous perturbations, which
will be a topic for study in the future.

\begin{acknowledgments}

 This work was supported by the National Natural
Science Foundation of China under Grants Nos. 10935013 and 11075083,
Zhejiang Provincial Natural Science Foundation of China under Grants
Nos. Z6100077 and R6110518, the FANEDD under Grant No. 200922, the
National Basic Research Program of China under Grant No.
2010CB832803, the NCET under Grant No. 09-0144, and K.C. Wong Magna
Fund in Ningbo University.

\end{acknowledgments}

\end{document}